# Legal Risks of Adversarial Machine Learning Research

Ram Shankar Siva Kumar*, *Microsoft, ramk@microsoft.com*
Jonathon Penney*, *Citizen Lab, University of Toronto, jon@citizenlab.ca*
Bruce Schneier, *Harvard Kennedy School, schneier@schneier.com*
Kendra Albert, *Harvard Law School, kalbert@law.harvard.edu*

## Introduction

Adversarial machine learning is the systematic study of how motivated adversaries can compromise the confidentiality, integrity, and availability of machine learning (ML) systems through targeted or blanket attacks. The problem of attacking ML systems is so prevalent that CERT, the federally funded research and development center tasked with studying attacks, issued a broad vulnerability note on how most ML classifiers are vulnerable to adversarial manipulation.[1] Corporations and governments are paying attention. Google,[2] IBM,[3] Facebook,[4] and Microsoft[5] have committed to investing in securing machine learning systems. The US is putting security and safety of AI systems as a top priority when defining AI regulation,[6] with the EU releasing a complete set of non-binding checklists as part of its Trustworthy AI initiative.[7]

Research in this field is booming. Since 2014, more than 2,000 papers have been posted in arXiv.[8] Researchers used adversarial ML techniques to identify flaws in Facebook's micro-targeting algorithm.[9] In an instance of perturbation attack, researchers were able to veer Tesla's self-driving system off the highway via simple stickers.[10] These kinds of attacks are particularly prevalent against defensive tools developed by the cybersecurity industry. For instance, researchers were able to trick the ML system powering Cylance, a commercial antivirus engine, into misrecognizing ransomware as benign software.[11] In another example, two security researchers bypassed Proofpoint, a commercial email protection, by first copying the underlying ML model and brute-forcing it offline, then launching an online attack crafting emails that escaped the system. This became the first vulnerability that was exploited on an ML system to be entered into the National Vulnerability database.[12]

---

[1] https://kb.cert.org/vuls/id/425163/ (noting "Machine learning classifiers trained via gradient descent are vulnerable to arbitrary misclassification attack")
[2] https://ai.google/responsibilities/responsible-ai-practices/?category=security ("Safety and security entails ensuring AI systems behave as intended, regardless of how attackers try to interfere.")
[3] https://www.ibm.com/cloud/architecture/architectures/securityArchitecture/watson-security
[4] https://spectrum.ieee.org/tech-talk/artificial-intelligence/machine-learning/facebook-ai-launches-its-deepfake-detection-challenge ("...Facebook's AI Red Team, which analyzes the threats that AI poses to the social media giant")
[5] https://docs.microsoft.com/en-us/security/engineering/securing-artificial-intelligence-machine-learning (noting "AI designers will always need to ensure the confidentiality, integrity and availability of sensitive data, that the AI system is free of known vulnerabilities, and provide controls for the protection, detection and response to malicious behavior against the system or the user's data")
[6] https://www.whitehouse.gov/wp-content/uploads/2020/01/Draft-OMB-Memo-on-Regulation-of-AI-1-7-19.pdf (noting "When evaluating or introducing AI policies, agencies should be mindful of any potential safety and security risks, as well as the risk of possible malicious deployment and use of AI applications")
[7] https://ec.europa.eu/digital-single-market/en/news/ethics-guidelines-trustworthy-ai (noting "AI systems need to be resilient and secure. They need to be safe, ensuring a fall back plan in case something goes wrong, as well as being accurate, reliable and reproducible. That is the only way to ensure that also unintentional harm can be minimized and prevented")
[8] https://nicholas.carlini.com/writing/2019/all-adversarial-example-papers.html (see txt file containing all papers)

\* Equal Contribution

This research is not without legal risks. Studying or testing the security of any operational system potentially runs afoul the Computer Fraud and Abuse Act (CFAA),[13] the primary United States federal statute that creates liability for hacking.[14] Originally enacted as a narrow law in 1984, it has subsequently been expanded, with new prohibitions added in the years since.[15] The broad scope of the CFAA has been heavily criticized, with security researchers among the most vocal.[16] They argue the CFAA--with its rigid requirements and heavy penalties--has a chilling effect on security research.[17] Adversarial ML security research is likely no different.

However, prior work on adversarial ML research and the CFAA is sparse. In a 2018 article, Ryan Calo et al. explored adversarial attacks under the CFAA but only examined three classes of adversarial ML attacks (evasion, poisoning, and data inversion).[18] Other relevant published commentary is similarly limited. For instance, one recent paper examined only evasion and poisoning attacks,[19] while another considered perturbation, extraction, and poisoning attacks but did not examine important inconsistencies in how courts have applied relevant CFAA provisions.[20] We do so here.

There are two goals for this paper. For legal practitioners, we describe the complex and confusing legal landscape of applying the CFAA to adversarial ML. For adversarial ML researchers, we describe the potential risks of conducting adversarial ML research. We also conclude with an analysis predicting how the US Supreme Court may resolve some present inconsistencies in the CFAA's application in *Van Buren v. United States*,[21] an appeal expected to be decided in 2021. We argue that the court is likely to adopt a narrow construction of the CFAA, and that this will actually lead to better adversarial ML security outcomes in the long term.

---

[9] Faizullabhoy, Irfan, and Aleksandra Korolova. "Facebook's advertising platform: New attack vectors and the need for interventions." *arXiv preprint arXiv:1803.10099* (2018).
[10] Xi, Bowei. "Adversarial machine learning for cybersecurity and computer vision: Current developments and challenges." *Wiley Interdisciplinary Reviews: Computational Statistics*: e1511.
[11] https://skylightcyber.com/2019/07/18/cylance-i-kill-you/
[12] https://nvd.nist.gov/vuln/detail/CVE-2019-20634
[13] 18 U.S.C. § 1030.
[14] Kossoff, Jeff. (2019). Cybersecurity Law (Wiley, 2019) at 172.
[15] Orin S. Kerr, *Vagueness Challenges to the Computer Fraud and Abuse Act*, 94 Minn. L. Rev. 1561, 1561 (2010); Kossoff 172-173.
[16] Kosseff 212-213.
[17] Kosseff 213.
[18] Calo, Ryan, et al. "Is Tricking a Robot Hacking?." University of Washington School of Law Research Paper 2018-05 (2018);
[19] Shankar Siva Kumar, Ram, et al. "Law and Adversarial Machine Learning." *arXiv preprint arXiv:1810.10731* (2018).
[20] Natalie Chyi. (2020). Examining the CFAA in the Context of Adversarial Machine Learning. https://www.legaltechcenter.net/a-i/commentary/

## CFAA and Adversarial ML

We consider two CFAA sections particularly relevant to adversarial machine learning. First, intentionally accessing a computer "without authorization" or in a way that "exceeds authorized access" and as a result obtains "any information" on a "protected computer" (section 1030(a)(2)(C)). Second, intentionally causing "damage" to a "protected computer" without authorization by "knowingly" transmitting a "program, information, code, or command" (section 1030(a)(5)(A)). These are both criminal prohibitions, but the CFAA also includes a private right of action. This allows any person to sue if they have incurred damages or losses due to a CFAA violation, including these two provisions.[22] These sections do not exhaust the potential liabilities under the CFAA, but do cover the most common adversarial ML attacks.

**Intentional Access Without or Exceeding Authorization -- Section 1030(a)(2)**
The CFAA has been inconsistently applied. Case law on its interpretation has been described as "fragmented"[23] and "unclear,"[24] with courts "expressing uncertainty and confusion".[25] One of the most contentious inconsistencies is section 1030(a)(2)(C)'s interpretation, which prohibits anyone who "intentionally accesses a computer without authorization" or "exceeds authorized access," and "thereby obtain information from any protected computer."[26] On a few points, courts have generally agreed. They have interpreted the term "protected computer" very broadly. For example, it includes any computer connected to the internet.[27] They have also generally found that "without authorization" means access without permission to do so. However, on the meaning of the term "exceeds authorized access," there are significant disagreements. The language applies to insiders: users that already have authorized access to a computer system -- such as a user sending query requests to an ML system -- but who do something to *exceed* that authorization in accessing any information on the computer.

What constitutes exceeding authorized access? There is presently a 4-3 split among the circuit courts of appeals in the United States. (For non-lawyers, the federal circuit courts cover particular regions of the country, and govern interpretation of the law by federal trial courts in those areas in the absence of a Supreme Court ruling.) The First, Fifth, Seventh, and Eleventh Circuits, which include Illinois, Massachusetts, and Texas, have adopted a broader interpretation, finding that "exceed authorized access" includes accessing information on a computer system for an "improper purpose," which usually means breaching some agreement, policy, or terms of service. Some examples have included accessing account information of another user that the accused was not managing or supervising. Another example would be a departing employee downloading company information contrary to confidentiality agreements or violating network use or access policies. For the purposes of our analysis below, we will refer to this as the "broader interpretation" of 1030(a)(2).

---

[21] Van Buren v. United States, 940 F.3d 1192 (11th Cir. 2019), petition for cert. filed (U.S. Dec. 18, 2019) (No. 19-783) (pending decision on appeal).
[22] Id. § 1030(g)
[23] Michael J. O'Connor, *The Common Law of Cyber-Trespass*, 85 Brook. L. Rev. 421, 422 (2020)
[24] Orin S. Kerr, *Vagueness Challenges to the Computer Fraud and Abuse Act*, 94 Minn. L. Rev. 1561, 1562 (2010).
[25] Orin S. Kerr, *Norms of Computer Trespass*, 116 Colum. L. Rev. 1143, 1145 (2016).
[26] 18 U.S.C. § 1030(a)(2)(C).
[27] *United States v. Kramer,* 631 F.3d 900, 902 (8th Cir. 2011) (noting the definition of "computer" is "exceedingly broad," and concluding an ordinary cell phone is a computer); *United States v. Nosal (Nosal II),* 844 F.3d 1024, 1050 (9th Cir. 2016), cert. denied, 138 S. Ct. 314 (2017) (noting "protected computers" include "effectively all computers with Internet access…")

By contrast, the Second, Fourth, and Ninth Circuit Courts of Appeal, which include New York and California, have adopted a narrow construction of the term. They have held that simply accessing information for an improper purpose, without more, does not constitute a section 1030(a)(2) violation. Rather, a person only "exceeds authorized access" if they are prohibited from accessing information for any reason. In practice, on this more narrow interpretation, a defendant "exceeds authorized access" when a user (i) had access to a computer system; (ii) their authorized access was limited to certain information on the system; and (iii) the user bypasses or circumvents a "technological access barrier" or "code-based" restriction to access additional information.[28] For the purposes of our analysis below, we will refer to this as the "narrow interpretation" of 1030(a)(2).

The remaining circuit courts have not ruled definitively on this question. So there are some regions in the country employing a narrow interpretation; some regions employing a broader interpretation; and some remaining regions where this interpretive issue has not yet been addressed, creating a confused and uncertain state of affairs for how the CFAA is applied across the country. The fragmented nature of the existing law is visualized in the following map (**Figure 1**):

---

[28] *See* Brenda Sharton et al., "Key Issues in Computer Fraud and Abuse Act (CFAA) Civil Litigation", *Practical Law (Thomson Reuters)* 1, 4 (2018) (summarizing the narrow reading); *United States v. Nosal I,* 676 F.3d 854, 863-864 (9th Cir. 2012) (en banc) ("his narrower interpretation is also a more sensible reading of the text and legislative history of a statute whose general purpose is to punish hacking — the circumvention of technological access barriers — not misappropriation of trade secrets"). Kosseff at 175 ("A narrow reading of the statute might lead to the conclusion that you only violate the CFAA if you commit a code-based violation").

**Figure 1: Differing CFAA Interpretation By Circuit Court Region**

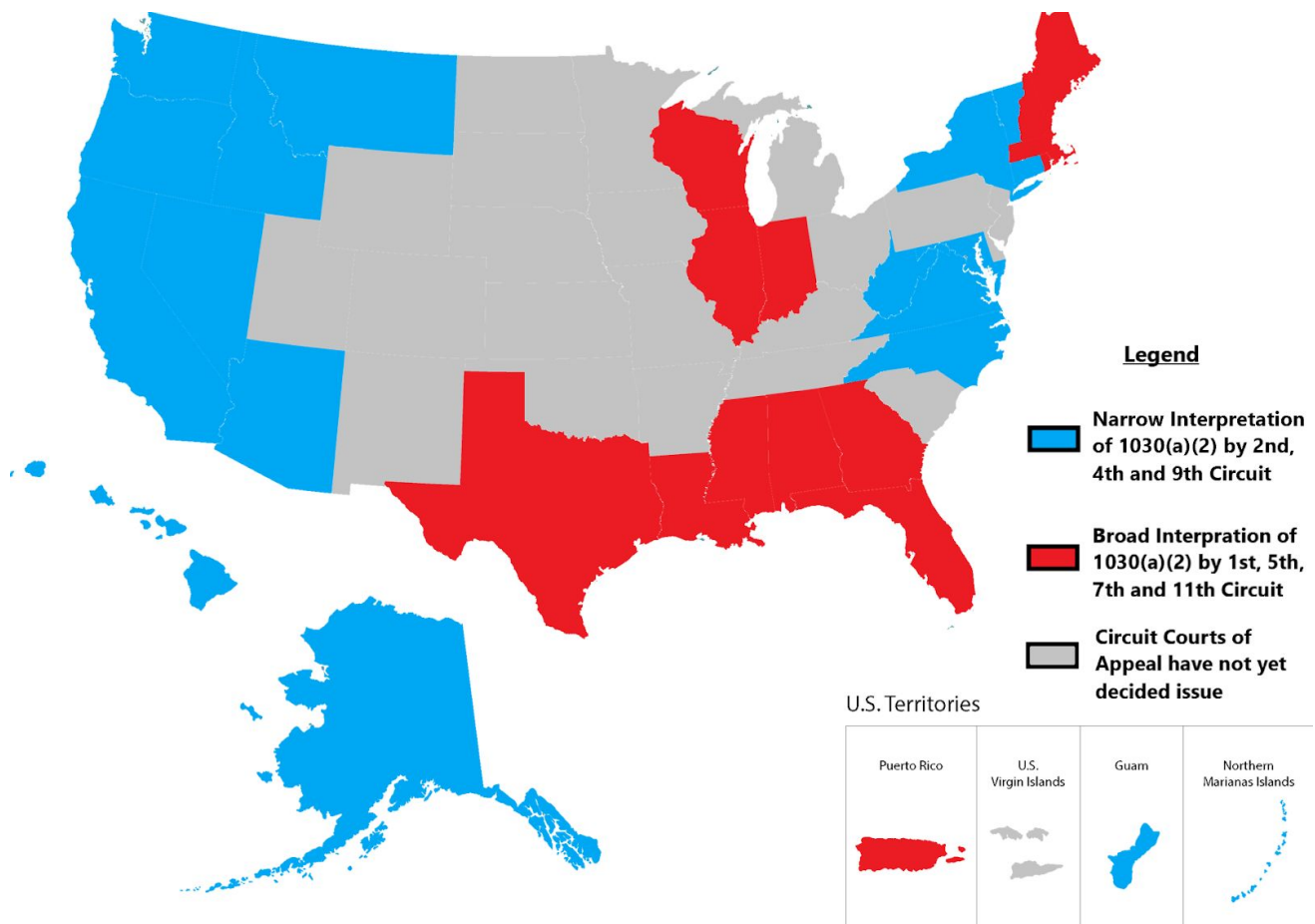

This confusing state of affairs may come to an end soon as the US Supreme Court has an opportunity to address these inconsistencies in *Van Buren v. United States*.[29] We discuss this case and its connections to adversarial ML in the final part of this paper.

**Intentional Access Without or Exceeding Authorization -- Section 1030(a)(5)**
Section 1030(a)(5)(A) prohibits anyone from "knowingly" transmitting a "program, information, code, or command" and thereby intentionally causing "damages" to a "protected computer." Unlike section 1030(a)(2)(C), this applies to outsiders and insiders equally. Though the CFAA does not define "transmission," courts have generally held that it can occur over the internet or through a physical medium, such as a USB stick.[30] For liability, a person must not just knowingly transmit, but also intentionally damage the protected computer. The CFAA defines "damage" as including "any impairment to the integrity or availability of data, a program, a system, or information."[31]

---

[29] Van Buren v. United States, 940 F.3d 1192 (11th Cir. 2019), petition for cert. filed (U.S. Dec. 18, 2019) (No. 19-783) (pending decision on appeal).
[30] Meridian Fin. Advisors, Ltd. v. Pence, 763 F. Supp. 2d 1046, 1061-62) (S.D. Ind. 2011)).
[31] § 1030(e)(8).

Denial-of-service (DOS) attacks -- where code or commands are sent over the internet or another medium, and overwhelm or disrupt the receiving system -- have also been prosecuted under this section. Other examples include sending malicious code, including trojans or viruses; code that deletes emails on the recipient's computer; code sent over a radio communications network used for first responder communications that intentionally interfere with those normal operations;[32] and bulk or mass emails that overwhelm the receiving system, disrupting normal business operations.[33]

# How the CFAA Affects Adversarial Machine Learning Research

Since its enactment, the CFAA has been applied to researchers who study operational systems. The exceptionally broad scope of both sections 1030(a)(2)(C) and 1030(a)(5) has implications for adversarial ML researchers -- and security researchers in general -- who may face criminal and civil liability for unauthorized access to a computer. Research on operational systems generally involves probing, or even breaking into, those systems. This includes those with permission to access a system as a legitimate user, whose access may be governed by terms of service. Merely testing in ways contrary to the terms of service, without anything more, could mean the researcher has exceeded authorized access and has violated the CFAA. The CFAA's severe penalties means such findings could have a chilling effect on security research.[34] Additionally, the circuit court split on the interpretation of section 1030(a)(2)(C) means that the extent of risk depends on the jurisdiction in which a CFAA claim may be adjudicated.

In this section, we illustrate how these CFAA provisions impact adversarial machine learning research more specifically. We assume a paid service that provides its users a machine learning service (say, image recognition). Users interact with this system by submitting a query, and obtaining the classification result (see **Figure 2**):

---

[32] United States v. Mitra, 405 F.3d 492 (7th Cir. 2005) (also finding a radio system is a computer).
[33] Pulte Homes, Inc. v. Laborers' Int'l Union of N. Am., 648 F.3d 295, 301 (6th Cir. 2011).
[34] https://www.theguardian.com/technology/2014/may/29/us-cybercrime-laws-security-researchers; https://www.eff.org/wp/protecting-security-researchers-rights-americas

**Figure 2: Black Box Setup**

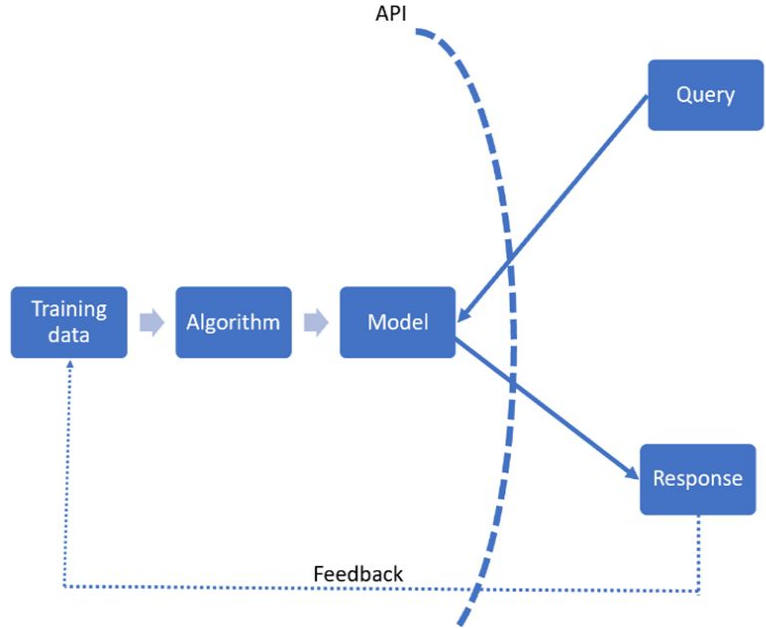

For our analysis, we assume that the attacker has no direct access to the training data, has no knowledge of the algorithm, and no knowledge about the features used in the algorithm. The attacker is also assumed to be a legitimate user of the system and their use of the system is governed by these rules, which are based on Google API's Terms of Service (TOS).[35]

Under those terms, users shall not:

1. Reverse engineer or attempt to extract the source code from any API or any related software, except to the extent that this restriction is expressly prohibited by applicable law
2. Interfere with or disrupt the APIs or the servers or networks providing the APIs
3. Scrape, build databases, or otherwise create permanent copies of such content, or keep cached copies longer than permitted by the cache header
4. Copy, translate, modify, create a derivative work of, sell, lease, lend, convey, distribute, publicly display, or sublicense to any third party
5. Misrepresent the source or ownership

Assuming this common setup, in the next section we consider a range of adversarial ML attacks in light of the CFAA, including taking into account the circuit split on how section 1030(a)(2)(C) should be interpreted. For clarity, though we often offer examples of each broad class of different adversarial ML attacks for illustrative purposes, our analysis aims to address how these CFAA provisions would apply to each broad class of different adversarial ML attacks, as opposed to a single instance of the attack. We also plot our analysis in **Figure 3**.

---

[35] https://developers.google.com/terms

**Exploratory Attacks: Attacks that only send queries and observe responses**

**Evasion Attack**
In evasion attacks, the attacker is able to modify the query to get any response that they desire, without bypassing the components behind the ML API. For instance, researchers were able to add specific noise to X-rays, which tricked an ML system into misclassifying abnormal X-rays as normal.[36] In the model evasion attack affecting Cylance antivirus engines, the researchers simply appended benign code from an online gaming repository to malicious ransomware, which caused the ML system to misclassify the code as benign software.[37]

Such attacks are unlikely to raise liability concerns under either section 1030(a)(2) or 1030(a)(5)(A). For the former provision, there is no section 1030(a)(2) violation on the narrow interpretation of "exceeds authorized access" because no technological access barrier or authentication gate is circumvented with the queries. The ML system is only tricked into providing an incorrect response. But there is also no violation under a broader interpretation, either, as this attack does not breach the TOS: no reverse engineering, scraping, or creation of a derivative work. Liability under section 1030(a)(5)(A) is also unlikely. Even assuming the query constitutes a transmitted "program, information, code, or command" there are no "damages" within the meaning of the term. There is no "impairment" to the "integrity" or "availability" of the system or any related data,[38] as there is no "destruction, corruption, or deletion of electronic files", "physical destruction" of the system, nor "any diminution in the completeness or usability of the data on a computer system."[39]

**Model Stealing**
With model stealing attacks, the adversary is able to create a copycat version of the underlying ML model by strategically querying the model and observing the response, thereby recreating the intellectual property. For instance, researchers were able to recreate deep neural nets hosted by ML as service providers, such as Microsoft, Face++, IBM, Google, and Clarifai, that were not only accurate (accuracy >80%), but also cheap to mount (cost less than $3).[40] One approach to model stealing is by using the existing ML API as an oracle; that is, the attacker makes sufficient queries over all the classes of the ML model, observes the response, and from the (query, response) pairs, learns a new function.[41]

---

[36] Paschali, Magdalini, et al. "Generalizability vs. robustness: investigating medical imaging networks using adversarial examples." *International Conference on Medical Image Computing and Computer-Assisted Intervention*. Springer, Cham, 2018.
[37] I.d. 11
[38] § 1030(e)(8).
[39] TriTeq Lock & Sec. LLC v. Innovative Secured Solutions, LLC, Civ. Action No. 10 CV 1304, 2012 WL 394229, at 6; Kossoff at 197-198.
[40] Yu, Honggang, et al. "Cloudleak: Large-scale deep learning models stealing through adversarial examples." *Proceedings of Network and Distributed Systems Security Symposium (NDSS)*. 2020.
[41] Tramèr, Florian, et al. "Stealing machine learning models via prediction apis." *25th {USENIX} Security Symposium ({USENIX} Security 16)*. 2016.

With respect to section 1030(a)(2), on a narrow interpretation there is likely no violation for this attack because no technological access barrier or code-based restriction is circumvented. Queries are merely sent and inferences made based on observed responses. However, there may be a violation on the broader interpretation of 1030(a)(2), which will turn on how broadly a court interprets the relevant TOS governing the ML system and the insider user. On one hand, it would seem unlikely that a court would find that information that is *inferred* from legitimate query responses could be prohibited by TOS and lead to CFAA liability. On the other, as with the TOS we assume here, such ML use restrictions and policies often prohibit reverse engineering or creating databases based on query responses, even legitimate ones. So, it is *possible* a court would find a model stealing attack as an attempt to "reverse engineer" the model's "source code," contrary to Clause 1 of the TOS, which, as noted, are fairly typical terms of use or service. It might also constitute creating a "derivative work," also contrary to TOS Clause 4.

However, no section 1030(a)(5)(A) violation is likely here because, as with evasion attacks, the "damages" requirement is not met. There is no "impairment" to the "integrity" or "availability" of the system or any related data.

**Model Inversion and Membership Inference**
In model inversion attacks, attackers infer sensitive information about the private training data. One paper showed how, with even a black box access to the ML model used in personalized medicine, attackers could recover the private genetic markers of patients who were part of the training data.[42] These kinds of attacks were later expanded to the ML API setting, wherein researchers showed how attackers could exploit the confidence intervals revealed in responses to reconstruct the features used in the private training data.[43] Such attacks raise serious concerns, given that training data usually contains privacy-sensitive information.[44]

In membership inference attacks, the adversary is able to ascertain if a data point was part of the training data by strategically querying and observing the response. For instance, in a model trained on hospital discharge data and hosted in ML APIs such as Google Prediction API, researchers showed how attackers could use this technique to reconstruct the private information of participants who were part of the training data (such as the procedure the patient underwent) through simple generic information such as participant's age and gender.

---

[42] Fredrikson, Matthew, et al. "Privacy in pharmacogenetics: An end-to-end case study of personalized warfarin dosing." *23rd {USENIX} Security Symposium ({USENIX} Security 14)*. 2014.

[43] Fredrikson, Matt, Somesh Jha, and Thomas Ristenpart. "Model inversion attacks that exploit confidence information and basic countermeasures." *Proceedings of the 22nd ACM SIGSAC Conference on Computer and Communications Security*. 2015.

[44] Zhang, Yuheng, et al. "The secret revealer: generative model-inversion attacks against deep neural networks." *Proceedings of the IEEE/CVF Conference on Computer Vision and Pattern Recognition*. 2020.

Figure 3: Chart Visualization Adversarial ML Legal Risks

| Attack | Description | 1030(a)(2) Violation? (Narrow Interpretation by Second, Fourth, and Ninth Courts) | 1030(a)(2) Violation? (Broad Interpretation by First, Fifth, Seventh, and Eleventh Circuit Courts) | 1030(a)(5)(A) violation |
|---|---|---|---|---|
| Evasion Attack | Attacker modifies the query to get appropriate response | No | No | No |
| Model Inversion | Attacker recovers the secret features used in the model by through careful queries | No | Possible | No |
| Membership Inference | Attacker can infer if given data record was part of the model's training dataset or not | No | Possible | No |
| Model Stealing | Attacker is able to recover the model by constructing careful queries | No | Possible | No |
| Reprogramming the ML System | Repurpose the ML system to perform an activity it was not programmed for | No | Yes | Yes |
| Poisoning Attack | Attacker contaminates the training phase of ML systems to get intended result | No | Possible | Yes |
| Attacking the ML Supply Chain | Attacker compromises the ML models as it is being downloaded for use | Yes | Yes | Possible |
| Exploit Software Dependencies | Attacker uses traditional software exploits like buffer overflow to confuse ML systems | Yes | Yes | Yes |

Again, no section 1030(a)(5)(A) violation is likely here because there is no "impairment" to the "integrity" or "availability" of the system. A section 1030(a)(2)(C) violation is also unlikely on the narrow interpretation because no technological access barrier or code-based restriction is circumvented, only inferences based on query responses. However, the broader interpretation is trickier. As with model stealing attacks, a broader interpretation of the CFAA will also rely on how broadly a court interprets the TOS. A court may conclude that this is access for an improper purpose, and thus exceeds authorized access, as it constitutes "reverse engineering" the model's "source code,", contrary to the TOS. A court might also find *inferring* information about training sets or membership constitutes a step involved in "building" a database, contemplated by the TOS prohibitions in Clause 3. If so, this would likely constitute a CFAA violation.

**Reprogramming the ML System**
In this attack, the adversary is able to force the model to perform a task that the creator did not intend to do so by sending special queries. For instance, researchers[45] demonstrated how ImageNet, a system used to classify images into different categories, was repurposed to count squares. The authors of the paper explain how this attack could lead to abuse of ML systems: for instance, an attacker might want to create spam accounts but be impeded by CAPTCHA images that need to be solved. Using the reprogramming attack, the attacker could repurpose the image recognition system used in cloud-hosted photo storages to solve CAPTCHAs and therefore remove the impediment.

Here, there would be no section 1030(a)(2) violation if narrowly interpreted because, again, no technological access barrier has been bypassed. A broader interpretation could lead to a different conclusion, as this could certainly constitute an interference or disruption of the ML system contrary to Clause 2 of the TOS, as the integrity of the ML system is compromised. This attack also raises liability concerns under section 1030(a)(5)(A), as the special queries also constitute "code" or "commands" that are knowingly and intentionally "transmitted" to the ML system. As well, the attack causes an "impairment" to the "integrity" or "availability" of the ML system. This would constitute a *prima facie* violation.

**Attacks that taint the training data:**

**Poisoning Attack**
In order to adapt to possible shifts in the underlying data distribution, ML models are often retrained on the outputs that were generated by the model itself and the associated feedback from the user. For instance, in a spam classification setting, every time a user provides feedback to the model if the mail was incorrectly classified, the response is sent to the underlying models and retrained. This feedback channel, wherein the response is folded back into the training data, can be exploited by attackers to poison the training data. The canonical example is Microsoft's Tay chatbot, which used the responses generated by Tay's interactions with users as training data. Within 24 hours of its release, Microsoft had to decommission it, since trolls from the internet exploited this feedback channel, resulting in Tay sharing racist comments and images.[46]

---
[45] Elsayed, Gamaleldin F., Ian Goodfellow, and Jascha Sohl-Dickstein. "Adversarial reprogramming of neural networks." *arXiv preprint arXiv:1806.11146* (2018).

On a narrow reading, there is likely no section 1030(a)(2) violation, as no technological access barrier or authentication gate is bypassed in the attack. However, the attack arguably interferes with or disrupts the ML system's operations, contrary to Clause 2 of the TOS, so there may be liability on a broader interpretation.

This attack likely constitutes a section 1030(a)(5)(A) violation, as the queries are likely "code" or "commands" knowingly "transmitted" to the ML system. And the commands, by corrupting and "poisoning" the ML system, are very likely a form of "damage" within the meaning of the section in that it impairs the "integrity" or "availability" of the ML system. This is a *prima facie* violation.

**Attacks on underlying environment:**

**Attack ML Supply Chain**
In these classes of attack, the attacker subverts the machine learning system by tampering with the source code, build processes, or update mechanisms. Currently, there is a trend in using pretrained models in NLP and computer vision wherein ML models trained on gigantic generic datasets and powerful computers are made available to the public for task-specific customization.[47] Researchers [48] showed that attackers could mount a man-in-the-middle attack as these models are being downloaded for use, or simply insert malicious code in the public repository where the models are hosted.

This attack raises liability concerns under both sections. If the attack involved compromising technological security or access barriers in order to compromise the ML supply chain, this would constitute access "without authorization" for outsider attackers, and for insiders, it would also "exceed authorized access" on either a broad or narrow interpretation. Bypassing code-based restrictions and compromising source code on a protected computer without permission or authorization are classic cases of computer hacking prohibited under section 1030(a)(2), whether on a narrow or broad reading.

A section 1030(a)(5)(A) violation is also probable, depending on how the attack is carried out. If a technological barrier was bypassed, then it is likely a "code" or "command" was knowingly "transmitted" to the ML system, and as a result of this action, the "integrity" or "availability" of the ML system, via its supply chain, is "impaired." This is also a *prima facie* violation.

---

[46] https://blogs.microsoft.com/blog/2016/03/25/learning-tays-introduction/ (noting that "Unfortunately, in the first 24 hours of coming online, a coordinated attack by a subset of people exploited a vulnerability in Tay. Although we had prepared for many types of abuses of the system, we had made a critical oversight for this specific attack. As a result, Tay tweeted wildly inappropriate and reprehensible words and images")
[47] https://huggingface.co/calculator/
[48] Gu, Tianyu, Brendan Dolan-Gavitt, and Siddharth Garg. "Badnets: Identifying vulnerabilities in the machine learning model supply chain." *arXiv preprint arXiv:1708.06733* (2017).

**Exploit Software Dependencies**
Here the adversary, instead of attacking the training data, model, or probing via queries, attacks the underlying infrastructure on which the ML system is built. For instance, researchers[49] showed how attackers could exploit the unpatched vulnerabilities in popular ML packages like numpy and tensorflow, and violate confidentiality, integrity, and availability guarantees.

This attack also likely leads to liability on both counts. Exploiting software vulnerabilities in the ML system would violate section 1030(a)(2) because it would mean technological access barriers are bypassed without authorization or exceeding authorization, causing damage or loss. Similar to our reasoning concerning attacks on the ML supply chain, either a narrow or broad interpretation leads to the same conclusion here.

Again, depending on the nature of the attack, there may be liability under section 1030(a)(5)(A) as well. If a technological barrier was bypassed using exploit code, then it is likely a "code" would have been knowingly "transmitted" to the ML system, ultimately impairing the ML system's "integrity" or "availability". This, too, would be a *prima facie* violation.

# *Van Buren* and Future Directions

Our analysis has attempted to navigate significant inconsistencies in how the CFAA has been applied. Some of these interpretive issues may be settled in the US Supreme Court's pending decision in *Van Buren v. United States*, which will probably be decided in 2021.[50] In *Van Buren*, the accused was a police officer who accessed a police database for an "improper purpose"; that is, not for any policing related work but to sell information from the database to a third party. The third party turned out to be part of an FBI sting operation, and the accused was convicted of a felony under section 1030(a)(2). The Eleventh Circuit upheld the conviction and the accused has now appealed to the Supreme Court. The Court must decide whether the accused, in accessing the database for improper purposes, "exceeded authorized access."

In short, the Supreme Court has the opportunity to resolve the circuit split over how to apply this language in section 1030(a)(2), endorsing either a narrow or broader interpretation, and creating greater certainty in the CFAA's application. The decision would likely also impact other interpretive issues under the CFAA, such as whether a mere TOS violation can constitute access "without authorization" for the purposes of other subsections in the anti-hacking statute.

How will *Van Buren* be decided? In our view, the Ninth Circuit's more narrow construction of the CFAA, as articulated in *Nosal I*, is more faithful to the CFAA's original and central focus as an anti-hacking statute. As the court in *Nosal I* noted, Congress enacted the CFAA "primarily to address the growing problem of computer hacking," and a narrow interpretation that focused on hacking and bypassing technological barriers would be more consistent with that focus "rather than turning [the CFAA] into a sweeping Internet-policing mandate."[51] A broader CFAA interpretation effectively criminalizes contractual breach, by turning breaches of TOS or other common private computer use policies, into criminal violations. This would mean, the court notes, that "millions of unsuspecting individuals would find that they are engaging in criminal conduct."[52] In other words, a narrow construction is more consistent with the CFAA's original aims and also avoids a troubling outcome wherein internet users are criminally liable for simple TOS violations that most people commit every day, knowingly or not.

---

[49] Xiao, Qixue, et al. "Security risks in deep learning implementations." *2018 IEEE Security and Privacy Workshops (SPW)*. IEEE, 2018.
[50] Van Buren v. United States, 940 F.3d 1192 (11th Cir. 2019), petition for cert. filed (U.S. Dec. 18, 2019) (No. 19-783) (pending decision on appeal).

If we are correct and the Supreme Court follows the Ninth Circuit's narrow construction, this will have important implications for adversarial ML research. In fact, we believe that this will lead to better security outcomes in the long term. First, if ML security researchers and industry actors cannot rely on expansive TOS to deter against certain forms of adversarial attacks, then this should provide a powerful incentive to develop more robust technological and code-based measures to protect against such attacks. And with a more narrow construction of the CFAA, ML security researchers will be less likely chilled from conducting tests and other exploratory work on ML systems, again leading to better security in the long term.

Second, even if TOS can be relied on, they are unlikely to deter truly bad actors. Average users do not read TOS, and more sophisticated adversaries are unlikely to be deterred from attacks simply by a clause in a computer use policy. As such, these policies and contractual measures provide little proactive protection against adversarial attacks, while often deterring legitimate researchers from either testing systems or reporting results. However, the actors *most* likely to be deterred are ML security researchers who *would* pay attention to TOS and may be chilled from research due to fear of CFAA liabilities.[53] On this angle of view, expansive terms of service may be a legalistic form of security theater: performative, providing little actual security protection, while actually chilling practices that may lead to better security.

Third, at most, a broad interpretation of the CFAA on this count may provide after-the-fact avenues for loss or damage recovery, but there are already existing legal options. Though remedies are certainly different, victims suffering damages or losses suffered due to a TOS violation can seek redress under state laws, such as civil claims for breach of contract or tortious interference, for example. In short, there are good and compelling reasons as to why the CFAA should remain a federal anti-hacking statute not a sweeping anti-competition statute. And this may, in the long run, lead to better adversarial ML security research and outcomes.

# Acknowledgement

We would like to thank Beth Friedman for her thoughtful feedback and edits to the paper.

---

[51] Nosal I at 859-862.
[52] Nosal I at 861.
[53] See for example: Sandvig v. Barr, No. CV 16-1368, 2020 WL 1494065 (D.D.C. 2020) (finding proposed academic research plans concerning consumer website practices--that would violate TOS of those consumer websites--would not lead to criminal liability under the CFAA).